\documentclass[preprint, oneside, onecolumn,10pt]{elsarticle}

\usepackage{etoolbox}
\usepackage[utf8]{inputenc}
\DeclareUnicodeCharacter{2032}{FIXME}

\usepackage{mathrsfs,mathtools,bbold}
\usepackage{amsmath,amssymb}
	\allowdisplaybreaks[1]
\usepackage{amsfonts}
\usepackage{float}
\usepackage[colorlinks,linktocpage]{hyperref}
\usepackage[dvipsnames]{xcolor}
\usepackage{physics}
\usepackage{slashed}
\usepackage{bm, cancel}
\usepackage[flushleft]{threeparttable}
\usepackage{multirow}
\usepackage{booktabs}
\usepackage{verbatim}

\usepackage[normalem]{ulem}

\hypersetup{colorlinks=true,linkcolor=blue,citecolor=blue,urlcolor=blue}

\biboptions{numbers,sort&compress}
\journal{\color{gray}\phantom{arxiv-hep}}
\begin{document}

\begin{frontmatter}

\title{Compact representation for electroweak lepton sector }

\author[label1]{P. I. Porshnev}
\ead{pporshnev@gmail.com}

\address[label1]{Past affiliation: Physics Department, Belarusian State University, Minsk, Belarus}

\begin{abstract}
A new representation of electroweak lepton sector is proposed. It consists of two Weyl spinors per one lepton family. It is shown that proposed representation is fully equivalent to the conventional left-handed iso-doublet. New type of plane wave solutions can be found under certain additional assumptions. 
\end{abstract}



\end{frontmatter}

\baselineskip=1.15\baselineskip

\section{Introduction}

Sometimes a new representation of well-established formalisms might lead to new insights or calculational benefits. The remarkable example is the spinor-helicity  method  \cite{dixon_brief_2013, elvang_scattering_2015} which greatly simplifies calculations of scattering amplitudes. 

In this work, we propose a new representation which combines two Weyl spinors, one for a left-handed charged lepton and one for its neutrino. Both our and Dirac representations are reducible, since they include two Weyl spinors each. Out of three Weyl spinors per lepton family, the Dirac one combines both electron spinors into one quantity, while we combine two SU2 spinors into one quantity $\psi$, exactly as the iso-doublet $L=(\psi_\nu, \psi_L)$ does. The only difference is we removed the trivial zeros from $L$; hence our construct $\psi$ is as artificial as the iso-doublet $L$ itself. In any way, these alternative combinations are equivalent to one another if proper mapping and corresponding representation-dependent operators are used.

We have been inspired to come up with this new representation by two sources. Specifically, quoting the first source - \cite[p. 704]{peskin_introduction_1995} - \emph{``Since the left- and right-handed fermions live in different representations of the fundamental gauge group, it is often useful to think of these components as distinct particles, which are mixed by the fermion mass terms''}. Similarly, modeling neutrinos as Majorana particles and quoting \cite{witten_lepton_2001} - \emph{``If lepton number is not conserved, one can
treat the left-handed neutrino and right-handed antineutrino as two different helicity states of one particle, and combine them to make a massive spin 1/2 particle''}. However, if the lepton number is not conserved, then the neutrino and antineutrino cannot be viewed as two different states of one particle \cite{witten_lepton_2001}.  Taking into account these ideas and extending them, we combine both left-handed electron and its right-handed antineutrino into our $\psi$, since they live in the iso-SU2 gauge representations ($L$ and $\bar L$), as opposite to their corresponding particles with opposite chiralities which live in iso-singlets. From this angle, two very different particles from the electroweak iso-doublets $L$ and $\bar L$, both left-handed electron and its right-handed antineutrino, must be somewhat ``closer'' to one another, than left- and right-handed electrons to one another \footnote{The second relevant quote from \cite[p. 704]{peskin_introduction_1995} says that ``The solution of this problem will reinforce the idea that the left- and right-handed fermion fields are fundamentally independent entities, mixed to form massive fermions by some subsidiary process.''}. This work intends to clarify what such a ``closeness'' might mean from the physical point of view. The practical benefit is also that this new representation is more compact compared to the conventional weak iso-doublets.  

Inspired by these insights, we rewrite the conventional left-handed part of EW formalism of massless leptons into an alternative, but fully equivalent, formalism by switching to the bispinor representation that combines two SU2 spinors of one lepton family. We show that no features of conventional EW model related to iso-doublets is lost. We then discuss how a new type of plane wave solutions can be obtained under certain additional assumptions. In this study, we focus only on the lepton sector. However, the proposed method  is really a gateway to constructing a broader framework which can be applied to quarks and bosons; possible extensions of this model will be reviewed in a future paper. The extension to quarks is straightforward but lengthier while the boson sector requires some additional assumptions; this is the reason they will not be covered here. We use the chiral representation of gamma matrices in this paper. The terminology and notations follow \cite{peskin_introduction_1995} and \cite{greiner_gauge_2009}.

\section{Compact EW formalism}

\subsection{Lagrangian and equations of motion}
In this subsection, we briefly state the conventional definitions which will be used later on.

The lepton-boson part of EW Lagrangian without right-handed particle states is given by 
\begin{equation}\label{chFields:eq3111}
	(\mathcal{L}_{int})^L = \frac{g}{2\sqrt{2}} (J_+W_{+}+J_-W_{-}) - eA J_{e}+\frac{g}{2\cos\theta}\Big(J_n - J_{e}\cos2\theta  \Big) Z \,,
\end{equation}
where the four currents are defined as
\begin{gather}
\begin{aligned}\label{chFields:eq3039}
	&J_+ &&= 2\bar{\psi}_{\nu} \gamma^\mu \psi_L\,,&&\qquad\quad J_e &&=  \bar{\psi}_L\gamma^\mu \psi_L\,, \\[1ex]
	&J_- &&= 2\bar{\psi}_{L} \gamma^\mu \psi_\nu\,,&&\qquad\quad J_n &&= \bar{\psi}_{\nu} \gamma^\mu \psi_{\nu}\,.
\end{aligned}
\end{gather}
Since the electromagnetic current $J_e=J^L_{em}$ in this work is always given by its left-handed part, we simplified its notation by dropping the superscript $L$. The first three currents couple only to their corresponding potentials, meaning that there seems to be one-to-one correspondence between three EW currents and three EW potentials. The $Z$-field however couples to both $J_n$ and $J_e$ with different but comparable strength ($\cos2\theta\approx 0.5$ where $\theta$ is the mixing angle). Two bispinors
 \footnote{Alternatively, a Dirac bispinor $\psi$ is called by a four-component Dirac spinor \cite[p. 6]{peskin_lectures_2017}.}, that are used in \eqref{chFields:eq3039},
  are given by
\begin{equation}\label{chFields:eq4000}
	\psi_L = \mqty(u_L\\0)\,,	\qquad\qquad \psi_\nu = \mqty(v_L\\0)\,,
\end{equation} 
They correspond to both massless charged lepton and neutrino that are described by two Weyl spinors $u_L$ and $v_L$ respectively. 
 
The covariant derivative for the massless left-handed doublet $L$ of electron and its neutrino is
\begin{equation}
	D_\mu L = (\partial_\mu  -\frac{i}{2}g' Y B_\mu- i g t \va{W}_\mu)L = (\partial_\mu  +\frac{i}{2}g'  B_\mu- i \frac{g}{2} \sigma_i W^i_\mu)L \,, 
\end{equation}
where the corresponding weak charge $t=1/2$ and hypercharge $Y=-1$. Expanding now the doublet $L$, we obtain
\begin{multline}
	D_\mu \mqty(\psi_\nu\\\psi_L) = \mqty(\partial_\mu\psi_\nu + \frac{i}{2}g'  B_\mu\psi_\nu  \\\partial_\mu\psi_L + \frac{i}{2}g'  B_\mu\psi_L)-  i \frac{g}{2}\mqty(W^3_\mu & W^1_\mu - iW^2_\mu\\ W^1_\mu +iW^2_\mu & - W^3_\mu)\mqty(\psi_\nu\\\psi_L)\\[1ex]
	=\mqty(\partial_\mu\psi_\nu + \frac{i}{2}g'  B_\mu\psi_\nu -i\frac{g}{2}  W^3_\mu\psi_\nu \\\partial_\mu\psi_L+ \frac{i}{2}g'  B_\mu\psi_L   +i\frac{g}{2}  W^3_\mu\psi_L)-   i \frac{g}{2}\mqty(W^+_\mu \psi_L\\ W^-_\mu \psi_\nu)\,. 
\end{multline}
where the last term mixes the electron and neutrino contributions. Using the orthogonal combinations of two neutral bosons $W^3_\mu$ and $B_\mu$, the covariant derivatives are re-cast as
\begin{multline*}
	D_\mu \psi_\nu  = \partial_\mu\psi_\nu + \frac{i}{2}\qty(g'  B_\mu -g  W^3_\mu)\psi_\nu -   i \frac{g}{2}W^+_\mu \psi_L \\[1ex]
	=\partial_\mu\psi_\nu -\frac{i}{2} \frac{g}{\cos\theta}Z_\mu \psi_\nu -   i \frac{g}{2}W^+_\mu \psi_L\,,
\end{multline*}\vspace{-5ex}
\begin{multline}\label{chFields:eq3329}
	D_\mu \psi_L  = \partial_\mu\psi_L + \frac{i}{2}\qty(g'  B_\mu +g  W^3_\mu)\psi_L -   i \frac{g}{2}W^-_\mu \psi_\nu \\[1ex]
	=\partial_\mu\psi_L + \frac{i}{2} \qty[2eA_\mu + \frac{g \cos 2\theta}{\cos\theta}Z_\mu]\psi_L -   i \frac{g}{2}W^-_\mu \psi_\nu\,.
\end{multline}
Therefore, two equations of motion are given by
\begin{gather}
\begin{aligned}\label{chFields:eq3042}
	&i\slashed D \psi_\nu &&= \qty(i\slashed \partial  +\frac{g}{2\cos\theta} \slashed Z) \psi_\nu +   \frac{g}{2}\slashed W^+ \psi_L&& =0\,,\\[1ex]
	&i\slashed D \psi_L &&= \qty(i\slashed \partial  - e\slashed A -\frac{g \cos 2\theta }{2\cos\theta} \slashed Z) \psi_L +   \frac{g}{2}\slashed W^- \psi_\nu&& =0\,,
\end{aligned}
\end{gather}
where we see again that the neutrino is not influenced by the EM potential while the EM term for electrons has the correct sign of electric charge.

\subsection{Redundancy} A single lepton family is described with two Dirac bispinors
in the Lagrangian \eqref{chFields:eq3111} and the equations of motion \eqref{chFields:eq3042}. However, since the Lagrangian $(\mathcal{L}_{int})^L$ includes only left-handed electrons $\psi_L$,  two complex-valued components per charged lepton are all that is really needed here. Now taking into account that massless neutrinos are also described by two-component spinor $v_L$
\begin{equation}\label{chFields:eq3036}
	\psi_L = P_L\psi=\mqty(u_L\\0)\,,	\qquad\qquad \psi_\nu = P_L\psi^\prime=\mqty(v_L\\0)\,,
\end{equation}
two general bispinors $\psi$ and $\psi^\prime$ have twice more degrees of freedom than what is really required in the conventional approach to describe the left-handed part of one lepton generation. Here, $P_{L/R}=(1\pm\gamma^5)/2$ are two chiral projectors. The use of Dirac bispinors in describing neutrinos is mostly for convenience \cite[p. 114]{berestetskii_quantum_1982}, though the accommodation of neutrino nonzero mass might change it; we ignore this complication for now. The most economical way is to directly use two-component spinors $(u_L, v_L)$ for all lepton types, see examples in \cite{peskin_lectures_2017} and \cite{dreiner_two-component_2010}. We will consider however an alternative approach.

Having used only the left-handed part in $P_L\psi$ to describe a charged lepton, we are left with two extra degrees of freedom in $\psi$ which otherwise would be associated with the right-handed charged lepton. Instead of introducing the second bispinor $\psi_\nu$, can we use these extra degrees of freedom in $\psi$ to describe neutrinos? Specifically, we would like to connect the neutrino wave function $\psi_\nu$ with a properly transformed bottom component of $\psi$
\begin{equation}
	\psi=\mqty(u_L\\u_R)\,,		\qquad\qquad \psi_\nu = U\,P_R \psi = \mqty(u^\prime_R\\0)\,,
\end{equation}
where $U$ is some transformation operator, and $u_R$ is the lower part of  bispinor $\psi$. 

It is important to clarify the following. No new physics which might stem from such an association is implied at this point. Instead, we focus on achieving the mathematical equivalency with the conventional left-handed part of EW theory while removing any redundancy from its description. Effectively, we search for a way to package two parts of EW iso-doublet into the single quantity $\psi$ which will be the direct sum of spinors $u_L$ and $v_L$ transformed in some way. The equivalency stems from the fact that such a procedure is fully reversible and is one-to-one.

Examples of transformations that swap left- and right-handed states are well known \cite[p. 44]{peskin_introduction_1995}. The Dirac bispinor representation is reducible that is clearly seen in the chiral gamma basis where the Lorentz generators are $2\cross2$ block-diagonal. A general matrix $U$ which swaps chirality states must then be of purely off-diagonal form in the chiral basis 
\begin{equation}
	U \sim \mqty(0&X\\Y&0) \qquad\to \qquad \begin{cases}
		\gamma^\mu\\
		\gamma^\mu\gamma^5
	\end{cases}\,,
\end{equation}  
where $X$ and $Y$ are some $2\cross 2$ matrices. There are eight basis matrices $(\gamma^\mu, \gamma^\mu\gamma^5)$ that are purely off-diagonal in the chiral basis. Since we are dealing with chiral states only and the matrix $\gamma^5$ is absorbed by chiral projectors
\begin{equation}
	\gamma^5 P_{R/L} = \pm P_{R/L}\,,
\end{equation}
we need to consider only $\gamma^\mu$ matrices. Therefore, a  chirality-swapping transformation can be taken in the following form
\begin{equation}\label{chFields:eq3037}
	\psi^\prime_{R/L}= U \psi_{L/R} = c^\mu\gamma_\mu \psi^*_{L/R}\,,
\end{equation}
where $c^\mu$ are some constant coefficients, and the complex conjugation might be also added. It is actually required, since the bottom spinor is associated with antiparticles which are right-handed. 

The charge conjugation transformation has exactly this form \cite[p. 96]{berestetskii_quantum_1982}
\begin{equation}
	\psi^C(x) = i\gamma^2 \psi^*(x)\,.
\end{equation}
Checking it out, we see that the neutrino spinor is obtained from $\psi$ as 
\begin{equation}\label{chFields:eq3041}
	\psi_\nu \sim P_L\psi^C = P_L i\gamma^2\mqty(u_L\\u_R)^*=P_L \mqty(\phantom{-}i \sigma_2 u^*_R\\-i \sigma_2u^*_L) \sim \mqty(v_L\\0)\,,
\end{equation}
if $v_L=i\sigma_2 u^*_R$ is associated with the transformed right-handed part of bispinor $\psi$.  The conventional iso-doublet $L$ is then replaced with the following bispinor
\begin{equation}\label{chFields:eq3341}
	L=\mqty(\psi_\nu\\\psi_L)\qquad\leftrightarrow\qquad \psi = \mqty(u_L\\i\sigma_2v_L^*)\,,
\end{equation}
where both of them have the equal number of nontrivial components. This representation is reversible and one-to-one, if the zero entries in $L$ are ignored. Accordingly, all four EW currents \eqref{chFields:eq3039} are unambiguously recovered as 
\begin{gather}
\begin{aligned}\label{chFields:eq3040}
	&J_+ &&=2\bar{\psi}_{\nu} \gamma^\mu \psi_L&&&&=\bar{\psi}^{C}\gamma^{\mu}\psi\,,\\[1ex]
	&J_- &&=2\bar{\psi}_{L} \gamma^\mu \psi_\nu&&&&=(\bar{\psi}^{C}\gamma^{\mu}\psi)^*\,,\\[1ex]
	&J_e &&=\bar{\psi}_L\gamma^\mu \psi_L&&=\frac{1}{2}(k^\mu - s^\mu)&&=\bar{\psi}\gamma^\mu P_L\psi \,,\\[1ex]
	&J_\nu &&=\bar{\psi}_{\nu} \gamma^\mu \psi_{\nu}&&=\frac{1}{2}(k^\mu + s^\mu)&&=\bar{\psi}\gamma^\mu P_R\psi \,,
\end{aligned}
\end{gather} 
where the bilinears $k_\mu=\bar\psi\gamma_\mu\psi$ and $s_\mu=\bar\psi\gamma_\mu\gamma^5\psi$ are defined as usual. The conventional EW currents on the left side are equal to the currents on the right side identically, component-by-component.  The electromagnetic and neutral currents are given as left- and right-handed parts of total current $k_\mu$ respectively if the representation \eqref{chFields:eq3341} is used. 
\\

\noindent\textbf{Transformations: } The key support for the compact formalism we develop here comes from the way such a combined quantity $\psi$ transforms  between inertial frames.  The boost and rotation transformations are given by block-diagonal matrices $S$ in the chiral representation of gammas
\begin{equation}
	S = \mqty(s_L &0\\0&s_R)\,,
\end{equation}
which means that shifting from one inertial frame to another does not mix the left-handed electron states with right-handed antineutrino ones
\begin{equation}
	\psi^\prime = \mqty(e^\prime\\\bar\nu^\prime)=\mqty(s_L &0\\0&s_R)\mqty(e\\\bar\nu)=\mqty(s_L e\\s_R \bar\nu)\,.
\end{equation}
This property is critical to avoid non-sensible results in the proposed framework.
\\

\noindent\textbf{New representation: } A word of caution should be shared. We have defined here the new representation $\psi$ which corresponds to the left-handed EW iso-doublet $L$. A problem might arise if one tries to apply the conventional operators, let us pick the charge conjugation operator \emph{C} for example, which is typically defined in the Dirac representation, to our construct $\psi$ which is defined in the different representation. Since any specific expression of \emph{C} is representation-dependent, such a definition must be used in a representation it is defined for. It should not be a problem, if one consistently uses the representation-dependent operators in their corresponding representations, since physical results must be representation-independent. Effectively, we have exploited the freedom in choosing a representation that is more convenient for the given task.   

The standard definition \emph{C} can still be applied to the chiral projections $P_L \psi$ where $\psi$ is defined in the new representation. The projection  $P_L \psi$  is the left-handed electron, which turns into the state with opposite charge upon applying the conventionally defined \emph{C}. Since our representation is one-to-one with two Weyl spinors, nothing is lost by combining two spinors into one $\psi$. The electron and neutrino Weyl spinors can always be extracted from our $\psi$ at any stage of calculations, and the standard charge conjugation operator can be applied to them individually. The price to pay for using our representation is that definitions for some operators now might look more cumbersome and include chiral projectors. However, there is always an identical mapping to the standard representation. It includes two Weyl spinors from our $\psi$ (left handed-electron and neutrino for SU2 iso-doublet), and one right-handed electron (SU2 iso-singlet) per one lepton family. We did not consider the latter ones in our approach which however does not conflict with right-handed electrons if terms with them are added to the Lagrangian.

Both our and Dirac representations are reducible, since they include two Weyl spinors each. Out of three Weyl spinors per lepton family, the Dirac one combines both electron spinors into one quantity, while we combine two SU2 spinors into one quantity, exactly as the iso-doublet $L=(\psi_\nu, \psi_L)$ does. The only difference is we removed the trivial zeros from $L$; hence this construct $\psi$ is as artificial as the iso-doublet $L$ itself. In any way, these alternative combinations are equivalent to one another if proper mapping and corresponding representation-dependent operators are used.

We have been inspired to come up with this new representation by the two sources \cite[p. 704]{peskin_introduction_1995} and \cite{witten_lepton_2001}, as it is discussed in the introduction in detail. It motivated us to combine both left-handed electron and its neutrino into our $\psi$, since they live in the same iso-SU2 gauge representation.  The benefit is that this new representation leads to new type of plane waves under certain assumptions.  
\\[-6pt]

Summarizing, the iso-doublet $L=(\psi_\nu, \psi_L)$  is replaced with the fully equivalent bispinor $\psi=(u_L,\,i\sigma_2v_L^*)$ which is  the direct sum of left-handed $u_L$ and right-handed $i\sigma_2v_L^*$ spinors respectively. At this point, no new physics has been introduced, even if the association of transformed left-handed neutrino with right-handed part of $\psi$ is suggestive. Simply, we compressed the left-handed iso-doublet $L$, half of which elements are zeros anyway, into the bispinor $\psi$ which does not have trivial entries. In doing so, nothing major has been lost or gained yet. Though, we might have gained in efficiency of describing the EW left-handed states by eliminating the redundant entries in iso-doublet $L$.

\subsection{Equation of motion and current conservation}\label{rqm:EqMotionSec}
Can two conventional equations of motion \eqref{chFields:eq3042}  be written as an evolution equation for the single bispinor $\psi$? Remember that in both equations \eqref{chFields:eq3042}, only the left-handed parts $\psi_\nu$ and $\psi_L$ participate. 

Substituting the definition \eqref{chFields:eq3041} into the first equation in \eqref{chFields:eq3042}, we obtain
\begin{equation}
	\qty(i\slashed \partial  +\frac{g}{2\cos\theta} \slashed Z) (P_L i\gamma^2\psi^*) +   \frac{g}{2}\slashed W^+ P_L\psi =0\,.
\end{equation}
Next, we complex conjugate it and multiply  with $i\gamma^2$
\begin{multline}
	i\gamma^2 \qty[-i (\gamma^\mu)^* \partial_\mu  +\frac{g}{2\cos\theta} (\gamma^\mu)^* Z_\mu] (P_L i\gamma^2\psi^*)^* +   i\gamma^2\frac{g}{2}(\gamma^\mu)^*  W^-_\mu (P_L)^*\psi^*\\[1ex]
	=\qty[i \gamma^\mu \partial_\mu  -\frac{g}{2\cos\theta} \gamma^\mu Z_\mu]i\gamma^2  (P_L i\gamma^2\psi) -  \frac{g}{2}\gamma^\mu   W^-_\mu i\gamma^2 P_L \psi^*\\[1ex]
	=\qty[i \gamma^\mu \partial_\mu  -\frac{g}{2\cos\theta} \gamma^\mu Z_\mu]P_R \psi -  \frac{g}{2}\gamma^\mu   W^-_\mu P_R i\gamma^2  \psi^*\,.
\end{multline}
Two motion equations are then given as
\begin{gather}
\begin{aligned}\label{chFields:eq3043}
	&i\slashed \partial (P_R\psi) -\frac{g}{2\cos\theta} \slashed Z (P_R\psi) -   \frac{g}{2}\slashed W^- P_R\psi^C&& =0\,,\\[1ex]
	&i\slashed \partial (P_L\psi) - e\slashed A (P_L\psi) -\frac{g \cos 2\theta }{2\cos\theta} \slashed Z (P_L\psi)+   \frac{g}{2}\slashed W^- P_L\psi^C&& =0\,.
\end{aligned}
\end{gather}
Now, two motion equations are re-written for the left- and right-handed parts of single bispinor $\psi$. We see that these two parts evolve differently under the EW forces, as expected. Let us next try to get rid of chiral projections of derivatives to obtain a single motion equation for $\psi$. 

For this purpose, we add the above equations together to obtain
\begin{multline}\label{chFields:eq3047}
	i\slashed \partial \psi - e\slashed A P_L\psi -\frac{g}{2\cos\theta} \slashed Z\qty( P_R +  \cos 2\theta P_L ) \psi -   \frac{g}{2}\slashed W^- \gamma^5\psi^C\\[1ex]
	= i\slashed \partial \psi -\frac{g}{2\cos\theta} \slashed Z\psi - \qty(e\slashed A -\frac{g\sin^2\theta}{\cos\theta}\slashed Z) P_L\psi-   \frac{g}{2}\slashed W^- \gamma^5\psi^C=0\,. 
\end{multline}
The last term makes this equation fundamentally different from the Dirac one. The neutral boson fields $A_\mu$ and $Z_\mu$ can be seen as influencing the particle momentum and spin which is similar to the regular Dirac equation, since they do not mix the upper and lower spinors. The last term which includes $\psi^C$ and charged EW bosons couples the left- and right-handed spinors. It is quite similar to the conventional mass term in this regard. Its role will be discussed in more detail in the next section.

What happens if we subtract one equation from another instead of adding them together? We obtain then from \eqref{chFields:eq3043}
\begin{multline}
	i\slashed \partial \gamma^5 \psi + e\slashed A P_L\psi -\frac{g}{2\cos\theta} \slashed Z\qty( P_R -  \cos 2\theta P_L ) \psi -   \frac{g}{2}\slashed W^- \psi^C\\[1ex]
	= \gamma^5\Big[-i\slashed \partial \psi + \gamma^5 e\slashed A P_L\psi -\frac{g}{2\cos\theta} \gamma^5\slashed Z\qty( P_R- \cos 2\theta P_L ) \psi -   \frac{g}{2}\gamma^5\slashed W^- \psi^C\Big]\\[1ex]
	= \gamma^5\Big[-i\slashed \partial \psi +  e\slashed A P_L\psi +\frac{g}{2\cos\theta} \slashed Z\qty( P_R+ \cos 2\theta P_L ) \psi +   \frac{g}{2}\slashed W^- \gamma^5\psi^C\Big]\\[1ex]
	= \gamma^5\Big[-i\slashed \partial \psi  +\frac{g}{2\cos\theta} \slashed Z \psi +\qty(e\slashed A -\frac{g\sin^2\theta}{\cos\theta}\slashed Z) P_L\psi +   \frac{g}{2}\slashed W^- \gamma^5\psi^C\Big]
	=0\,,
\end{multline}
which is identical to \eqref{chFields:eq3047} since $\gamma^5$ is non-singular.

The equation \eqref{chFields:eq3047} for $\psi$ can be obtained from \eqref{chFields:eq3043} in yet another way. Moving the chiral projectors in all terms to the left, we obtain
\begin{gather}
\begin{aligned}
	&P_L\Big(i\slashed \partial \psi -\frac{g}{2\cos\theta} \slashed Z \psi -   \frac{g}{2}\slashed W^- \psi^C\Big)&& =0\,,\\[1ex]
	&P_R\Big(i\slashed \partial \psi - e\slashed A \psi -\frac{g \cos 2\theta }{2\cos\theta} \slashed Z \psi+   \frac{g}{2}\slashed W^- \psi^C\Big)&& =0\,.
\end{aligned}
\end{gather}
Now taking into account the orthogonality of chiral projectors, we can make the round brackets identical
\begin{gather}
\begin{aligned}\label{chFields:eq3069}
	&P_L\Big(i\slashed \partial \psi - eP_R\slashed A \psi-[P_L +\cos 2\theta P_R]  \frac{g \slashed Z \psi}{2\cos\theta} -   \frac{g}{2}[P_L-P_R]\slashed W^- \psi^C\Big)&& =0\,,\\[1ex]
	&P_R\Big(i\slashed \partial \psi - eP_R\slashed A \psi -[P_L +\cos 2\theta P_R]  \frac{g \slashed Z \psi}{2\cos\theta}-   \frac{g}{2}[P_L-P_R]\slashed W^- \psi^C\Big)&& =0\,.
\end{aligned}
\end{gather}
Hence, the neutrino equation is obtained by the left projection of common equation while the electron one is obtained with the right projection. It does not contradict with the previous statement that the electrons and neutrinos are the left- and right-handed projections of $\psi$ respectively. In all terms of both motion equations \eqref{chFields:eq3069}, the wave function is protected with $\gamma^\mu$ from the left side; the projectors are flipped if moved across $\gamma^\mu$. This consideration is helpful to avoid a possible confusion.
\\

\noindent\textbf{Current conservation: } Remarkably, the equation \eqref{chFields:eq3047} satisfies the current conservation in exactly the same way as the regular Dirac equation does. Taking \eqref{chFields:eq3047} and its conjugate version followed by multiplication with $\bar\psi$ and $\psi$ respectively
\begin{gather}
\begin{aligned}
	&i\bar\psi\gamma^\mu(\partial_\mu \psi) -\frac{g}{2\cos\theta} \bar\psi\slashed Z\psi - \bar\psi\qty(e\slashed A -\frac{g\sin^2\theta}{\cos\theta}\slashed Z) P_L\psi-   \frac{g}{2}\bar\psi\slashed W^- \gamma^5\psi^C&&=0\,,\\[1ex]
	&i(\partial^\mu \bar\psi)\gamma^\mu\psi +\frac{g}{2\cos\theta} \bar\psi\slashed Z\psi + \bar\psi P_R\qty(e\slashed A -\frac{g\sin^2\theta}{\cos\theta}\slashed Z)\psi +   \frac{g}{2}\bar \psi^C\gamma^5\slashed W^+\psi &&=0\,,
\end{aligned}
\end{gather}
and then adding them together, we obtain 
\begin{equation}
	i\partial_\mu(\bar\psi\gamma^\mu\psi) - \frac{g}{2} \qty(W^-_\mu\bar\psi\gamma^\mu \gamma^5\psi^C-W^+_\mu\bar\psi\gamma^5\gamma^\mu \psi^C  )=i\partial_\mu(\bar\psi\gamma^\mu\psi) = 0\,.
\end{equation}
The only difference with the Dirac case is that we had to use the following identities
\begin{equation}
	P_R \gamma^\mu=\gamma^\mu P_L\,		\qquad\qquad\bar\psi\gamma^\mu \gamma^5\psi^C=\bar\psi^C\gamma^5\gamma^\mu \psi=0\,,
\end{equation}
which are universal in the sense that they do not depend on the representation of gammas,  the EW potentials,  and are valid for an arbitrary bispinor $\psi$. Therefore, the current conservation strictly follows from the motion equation \eqref{chFields:eq3047} without any additional constraints or assumptions.
\\

\noindent\textbf{Summary: } We have showed that the left-handed part of EW theory, which is defined by using two two-component spinors $u_L\sim\psi_L$ and $v_L\sim\psi_\nu$, can be re-written by using the single four-component bispinor $\psi=(u_L, u_R)$, if we assign $v_L = i \sigma_2 u_R^*$. Two conventional equations of motion \eqref{chFields:eq3042} are then translated into the single equation \eqref{chFields:eq3047} which is convenient to give in the following form
\begin{equation}\label{chFields:eq3344}
	i\slashed \partial \psi -\frac{g}{2\cos\theta} \slashed Z \psi_R - \qty(e\slashed A +\frac{g\cos2\theta}{2\cos\theta}\slashed Z) \psi_L-   \frac{g}{2}\slashed W^- \gamma^5\psi^C=0\,.
\end{equation}
It is fully equivalent to the left-handed part of conventional EW formalism for leptons. This single equation describes the evolution of both electron and neutrino spinors. Its left-handed chiral projection gives the equation of motion for transformed neutrino spinor $i\sigma_2 v_L^*$, while the right-handed projection describes the evolution of electron spinor $u_L$.

\section{Plane waves}\label{subSecLepton}
The conventional (and trivial) plane waves solutions are obtained from equation \eqref{chFields:eq3344} if the electroweak potentials $Z_\mu$, $A_\mu$, and $W_\mu$ are set to zero. The equation \eqref{chFields:eq3344} then becomes $i\slashed\partial \psi=0$ which solution is two Weyl plane waves that are independent from each other; they describes the left-handed election and its right-handed anti-neutrino  correspondingly in the case of \eqref{chFields:eq3344}. It is, of course, in the strict agreement with the conventional approach, as it is shown in the previous section.

Less trivial solutions in the form of plane waves can be obtained from \eqref{chFields:eq3344} if we assume that the electroweak potentials are nonzero even for free-moving leptons. A charged particle carries its Coulomb field (and corresponding charge) even in the absence of external fields. The conventional theory is somewhat controversial here. From the one hand, the conventional plane waves that are used in evaluating invariant amplitudes are obtained as solutions of Dirac or Weyl equations by setting electromagnetic fields to zero. So, they are effectively taken  as finite ones for free-moving particles. From the other hand, the Coulomb or Uehling potentials that are carried by free particles are infinite at $r\to 0$. Choosing the middle ground, we assume instead that the fields of free-moving leptons are finite and do not necessarily vanish, as it is taken in the conventional approach. In the approximation of free-propagating plane waves that we use here, finite values of electroweak potentials  ($A_0$, $Z_0$, and $W_0$) at location of bare charge are just a set of numbers, with no space-time dependence \footnote{Later, the superscript $self$ will be added to these potentials to indicate that they are viewed as potentials of free-moving particles.}.  We will not speculate here what form a more fundamental theory (with less infinities and leading to finite self-potentials) might have; all we need for our phenomenological approach to proceed is the assumption that  finite values of four electroweak potentials at bare charge origin exist and that they are not necessarily equal to zero.

One interesting class of solutions of \eqref{chFields:eq3344} can be obtained under several additional constraints. First, even if two left-handed leptons from the same lepton family are combined into iso-doublets, physically they propagate as separate and distinct particles, per the conventional view. This class of solution is obtained from \eqref{chFields:eq3344} by setting the fields to zero. Second, in the compact EW formalism, we propose here, the iso-doublets are replaced with single bispinors which left- and right-handed parts are associated with charged leptons and corresponding antineutrinos respectively. Extending the conventional case, we can attempt to find solutions as superpositions of upper and bottom spinors 
\begin{equation}\label{chFields:eq3337}
	\psi = \sqrt{m_e}\mqty(\chi\\0) + \delta \sqrt{m_\nu}\mqty(0\\\xi)\,,
\end{equation} 
where $m_e$ and $m_\nu$ are electron and neutrino masses respectively, $\chi$ and $\xi$ are two arbitrary spinors, and $\delta$ is some small parameter. The linear composition \eqref{chFields:eq3337} is given in the rest frame of both electron and antineutrino. It assumes that a hypothetical lepton state, which is described by \eqref{chFields:eq3337}, is left-handed at rest with an infinitesimally small amount of right-handed antineutrino component. Saying otherwise, the state \eqref{chFields:eq3337} is predominantly left-handed and negatively charged at rest. We have already showed in subsection \ref{rqm:EqMotionSec} that the current which is originated by this combined quantity $\psi$ is conserved absolutely, so there should not be any concern related to charge conservation.

The form \eqref{chFields:eq3337} is predominantly left-handed in the rest frame; it can however acquire an arbitrarily chirality under boosts. The unique feature of our represention is the direct connection between the state chirality and its charge which is given by the ratio of its upper and lower components. Boosting \eqref{chFields:eq3337} in the $z$-direction, we obtain
\begin{equation}\label{chFields:eq3338}
	\psi^\prime = \mqty(e^{\frac{\eta}{2}} P_d + e^{-\frac{\eta}{2}} P_u&0\\0&e^{\frac{\eta}{2} }P_u + e^{-\frac{\eta}{2}} P_d&0)\mqty(\sqrt{m_e} \chi\\ \delta\sqrt{m_\nu} \xi)\,,
\end{equation}    
where $P_{u/d}= (1\pm \sigma_3)/2$ are the spin $z$-projection projectors. Now, with the upper spinor in spin-down state $\chi=\smqty(0\\1)$, the upper component remains dominant if the rapidity $\eta\to \infty$ independently of the neutrino spin orientation. However, if both upper and lower components  are spin-up, then the left-handed charged component will become smaller than the right-handed neutral one at sufficiently high rapidity. Since the spin projection can also be changed by rotations, this behavior means that the superposition \eqref{chFields:eq3337} does not have invariant values of chirality or charge. 

It is however possible to apply further refinements to \eqref{chFields:eq3337} to ensure that its degree of chiral polarization and charge will remain unchanged under boosts and rotations. Let us consider a general boost $S(\Lambda)$ in the chiral representation of gammas. We do not need to consider rotations, since they do not change magnitudes of both (dotted and undotted) spinors. The bispinor $\psi$ changes under the boost with rapidity $\eta$ in direction $\va{m}$ as
\begin{multline}\label{chFields:eq3339}
	\psi^\prime = S(\Lambda)\mqty(\chi\\\xi)=\mqty(\cosh\frac{\eta}{2} - \sinh\frac{\eta}{2}\,\va{m}\cdot \bm{\sigma} &0\\0&\cosh\frac{\eta}{2} + \sinh\frac{\eta}{2}\,\va{m}\cdot \bm{\sigma})\mqty(\chi\\\xi)\\[1ex]
	=\mqty(\chi_1\cosh\frac{\eta}{2} -\sinh\frac{\eta}{2}(\chi_1 m_3 +\chi_2 m_1- i \chi_2 m_2)\\
	 \chi_2\cosh\frac{\eta}{2} +\sinh\frac{\eta}{2}(\chi_2 m_3 -\chi_1 m_1- i \chi_1 m_2)\\\xi_1\cosh\frac{\eta}{2} +\sinh\frac{\eta}{2}(\xi_1 m_3 +\xi_2 m_1- i \xi_2 m_2)\\\xi_2\cosh\frac{\eta}{2} -\sinh\frac{\eta}{2}(\xi_2 m_3 +\xi_1 m_1+ i \xi_1 m_2))\,.
\end{multline} 
This representation is given in the frame where the bispinor components are given by $\chi_i$ and $\xi_i$ respectively. We must find certain spinor polarizations that do not change the chirality of $\psi$ under boosts. Let us assume that in the given frame both $\chi_2=\xi_1=0$ which turns \eqref{chFields:eq3339} into
\begin{equation}\label{chFields:eq3340}
	\psi^\prime = \mqty(\chi_1(\cosh\frac{\eta}{2} -\sinh\frac{\eta}{2} m_3) \\
	 0\\0\\\xi_2(\cosh\frac{\eta}{2} -\sinh\frac{\eta}{2} m_3) )\,,
\end{equation} 
where we also set $m_1=m_2=0$.  It is then immediately clear that such a bispinor will not change its degree of chirality (the ratio of magnitudes of the upper spinor to the bottom one)  independently of either rapidity or boost direction.  While the expression \eqref{chFields:eq3340} is given for the boost in $z$-direction, the fact that magnitudes of boosted upper and bottom spinors will not change relative to one another under a general boost can be easily from \eqref{chFields:eq3339}.  Saying otherwise, if a bispinor is given by such a form in one frame and it is pre-dominantly left- or right-handed ($\abs{\chi_1}\gg\abs{\xi_2}$ or $\abs{\chi_1}\ll\abs{\xi_2}$ ), it will not change its chirality polarization under boosts or rotations. Ditto for the case  $\chi_1=\xi_2=0$.

The split of any bispinor (it is equivalent to the iso-doublet EW in our representation) into these two configurations of spinor components is given by the rank-two projectors $S_{\pm} = (1\pm \gamma^0\gamma^3)/2$ which filter spinor states in the chiral representation as
\begin{align}\label{chFields:eq3342}
	\psi_{s_z=-1/2}=S_{-}\psi = \smqty(0\\\chi_2\\\xi_1\\0)\,,	&\qquad\psi_{s_z=+1/2}=S_{+}\psi = \smqty(\chi_1\\0\\0\\\xi_2)\,.			
\end{align}
Therefore, the pure lepton states $\psi_l$ are obtained  by applying the spin projector to the iso-doublet $\psi$
\begin{equation}
	\psi_l = S_{\pm}  \psi\,.
\end{equation}
Depending on which spinor component is dominant, it could be associated with  a charged lepton-like or its corresponding antineutrino-like states.

Summarizing, new class of plane waves in the proposed framework is given as the fully spin-  and predominantly chirality-polarized state of the EW  iso-doublet in some chosen frame; it is convenient to choose the rest frame for such a purpose to stay consistent with the conventional way of particle classification.
\\

\noindent\textbf{Currents at rest: } If a bispinor $\psi$ is given by the forms \eqref{chFields:eq3342} at rest, the four EW currents \eqref{chFields:eq3040} that can be obtained from single $\psi$ are evaluated as
\begin{gather}
\begin{aligned}\label{chFields:eq3343}
	&J_+ &&=2\bar{\psi}_{\nu} \gamma^\mu \psi_L&&=\bar{\psi}^{C}\gamma^{\mu}\psi&&=2(i \chi\sigma_2\xi) \, V\,,\\[1ex]
	&J_- &&=2\bar{\psi}_{L} \gamma^\mu \psi_\nu&&=(\bar{\psi}^{C}\gamma^{\mu}\psi)^*&&=2(i \chi\sigma_2\xi)^* \, V \,,\\[1ex]
	&J_e &&=\bar{\psi}_L\gamma^\mu \psi_L&&=\bar{\psi}\gamma^\mu P_L\psi &&=(\abs{\xi}^2+\abs{\chi}^2) \, V\,,\\[1ex]
	&J_\nu &&=\bar{\psi}_{\nu} \gamma^\mu \psi_{\nu}&&=\bar{\psi}\gamma^\mu P_R\psi&&=(\abs{\xi}^2-\abs{\chi}^2) \, V\,,
\end{aligned}
\end{gather} 
where $V^\mu = (1,0,0,-1)$ is the light-like vector. Hence, all four EW currents that are generated by a single free lepton state $\psi$ are parallel and light-like.  This consideration is important for two reasons at least.

First, the four-momentum of massive lepton at rest is given as $p_\mu=(p_0, 0,0,0)$. Therefore, a free lepton-like state in our framework at rest ends up with only two independent four-vectors: the energy momentum $p_\mu$ and the spin vector $s_\mu=(0,0,0,s_3)$ which can be obtained as a linear combination of $p_\mu$ with any current from \eqref{chFields:eq3343}. Keep in mind however that by choosing forms \eqref{chFields:eq3342} we chose the rest frame with spin directed along $z$-axis.  No any other independent and non-null vectors can be obtained in this case. It is in the full agreement with observations that massive fermions at rest possess only energy-momentum and spin. If the EW currents in \eqref{chFields:eq3343} would be pointing in different directions, more independent and non-null vectors can be obtained for a free lepton which would lead to the clear contradiction with the conventional theory and experiment.

Second, it places certain restrictions on possible forms of self-induced potentials $A^{self} $, $Z^{self} $, or $W_{\pm}^{self} $. As we discussed before and in more detail in the Appendix, we do not assume that the EW potentials of free-moving leptons at charge origins are describable by the Maxwell or Klein-Gordon equations. Instead, we assume that in the framework of free-propagating plane waves, these self-potentials are either proportional to particle momentum $p_\mu$ or particle spin $s_\mu$, or some combination of these two vectors.  Therefore, potentials of free-moving leptons, let us pick $Z_\mu^{self} $, could be given as 
\begin{equation}\label{chFields:eq3345}
	Z_\mu^{self}  = (Z_0, 0, 0, Z_3)\,,
\end{equation}
under the requirement that $\abs{Z_0}\ne \abs{Z_3}$. Keep in mind again that by choosing forms \eqref{chFields:eq3342} we chose the rest frame with spin directed along $z$-axis. 
\hspace{-5pt}\footnote[7]{Having filtered spin states as in \eqref{chFields:eq3342} once, we can move into any other inertial frame by boosts and rotations.}
For example, the electromagnetic potential $A_\mu$ of electron at rest is traditionally given as $(A_0, 0, 0, 0)$ where $A_0$ is the electric potential; it might still have a small $A_3$ component which is related to electron spin or induced magnetic field. The similar consideration applies to both $Z^{self} $ and $W^{self} _\pm$.

\subsection{EW plane waves}
Assuming that self-action fields $A^{self} $, $Z^{self} $, and $W_{\pm}^{self} $ are originated and carried by a free-moving lepton, the corresponding motion equation  \eqref{chFields:eq3344} can be given as 
\begin{equation}\label{chFields:eq3073}
	\slashed p \psi -\frac{g}{2\cos\theta} \slashed Z^{self}\psi_R - \qty(e\slashed A^{self} +\frac{g\cos2\theta}{2\cos\theta}\slashed Z^{self}) \psi_L-   \frac{g}{2}e^{2ipx}\slashed W_-^{self}(x) \gamma^5\psi^C=0\,, 
\end{equation}
where the plane wave ansatz $\psi(x)=\psi e^{-ipx}$ was used; here $p_\mu$ is the phase momentum to be found. The last term is explicitly phase-dependent while all other terms are phase-independent. 

For consistency, we must request that the phase factor is canceled out in the combination  $e^{2ipx}W_-^{self}(x)$. We cannot eliminate it by gauge-transforming three other gauge potentials, since then an extra term will be generated by the derivative. It is the clear indication that the phases of charged self-fields $W_{\pm}^{self}(x)$ must be correlated with the phases of corresponding charged currents \eqref{chFields:eq3343} which are also phase-dependent. It must be $J_+(x)$ in the case of $W_-^{self}(x)$, since then its phase dependence offsets the phase factor in \eqref{chFields:eq3073}
\begin{equation}\label{chFields:eq3152}
	J_+(x) =\bar{\psi}^{C}(x)\gamma^{\mu}\psi(x) = e^{-2ipx} \underbrace{\bar{\psi}^{C}\gamma^{\mu}\psi}_{J_+}\,.
\end{equation}
For the purposes of this study, we do not need to specify a functional dependence between the self-potentials $W_{\pm}^{self}(x)$ and corresponding charged currents $J_{\mp}(x)$. All is needed are the phase correlation between the charged potentials and charged currents, and the form \eqref{chFields:eq3345} that defines the self-potentials in the chosen rest frame. 
A quick comment: after canceling the phase dependence (x-dependence in this case), the potential $W_-$ is assumed to be spacetime-independent in the manipulations below.  
\\

\noindent\textbf{Expectations: } As we discussed at the end of section \ref{rqm:EqMotionSec}, the equation \eqref{chFields:eq3047} or \eqref{chFields:eq3073} is defined for the quantity $\psi=(u_L, i\sigma_2 v_L^*)$.  Remember that its two chiral projections $P_L \psi = (u_L,0)$ and $P_R \psi = (0,i\sigma_2 v_L^*)$  correspond to conventional left-handed charged lepton and its antineutrino respectively. If we accept that these chiral projections can propagate independently of each other, we end up with exactly the conventional case, with no new physics. It is straightforward to demonstrate that  \eqref{chFields:eq3047} or \eqref{chFields:eq3073} describes the motion of two parts of EW iso-doublet in the exact correspondence with the conventional case. 

Instead, we can accept the view that two parts of EW iso-doublet might not be always separable in which case  the iso-doublet (which is represented by $\psi$ in our framework) propagates as a whole entity, with both nonzero upper and lower spinors. However, to match the reality, it must then have two modes:  neutrino-like and charged lepton-like. In solving the plane wave equation \eqref{chFields:eq3073}, two types of solutions are distinguished by having either $p^2=0$ or $p^2=m^2$ respectively for the first and second modes. 

For the second (massive) mode, both upper and lower spinors in $\psi=(\chi, \xi)$ must be nonzero and have the form \eqref{chFields:eq3342} at rest.  The  mass or inertia is generated by the lepton-boson interaction terms that are included into \eqref{chFields:eq3073}. The key expectation is that the magnitude of upper spinor is much larger than the magnitude of lower one ($\abs{\chi}\gg \abs{\xi})$.  For the neutrino-like mode, with zero or near zero masses, we expect that $\abs{\chi}\ll\abs{\xi}$. The corresponding eigenvalue for $p^2$ must either  be zero or be near zero. Our framework can easily accommodate nonzero neutrino masses, as it will be shown below.


\subsection{Solutions} 
The equation \eqref{chFields:eq3073} is the system of algebraic linear equations for components of $\psi$. Finding a general solution is quite a daunting task, since it depends on four real four-vector coefficients. Compare it with the conventional Dirac case of plane waves for which the general form depends only on $p_\mu$ and a unit spinor. We will start with finding solutions in the rest frame first.

Having eliminated the phase factor from  \eqref{chFields:eq3073} for the plane wave motion, it is convenient to introduce the following notations
\begin{equation}\label{chFields:eq3272}
	M\psi = \slashed p \psi -\slashed c\psi_R - \slashed b \psi_L-   \slashed d \gamma^5\psi^C=0\,,
\end{equation}
where the vector coefficients are defined as
\begin{gather}
\begin{aligned}\label{chFields:eq3273}
	&c_\mu &&= \frac{g}{2\cos\theta} Z_\mu^{self}\,,\\[1ex]
	&b_\mu &&= e A_\mu^{self} +\frac{g\cos2\theta}{2\cos\theta} Z_\mu^{self} \,,\\[1ex]
	&d_\mu &&=\frac{g}{2} W_-^{self}\,,
\end{aligned}
\end{gather}
here both $c_\mu$ and $b_\mu$ are real while $d_\mu$ is complex-valued. Since the self-induced potentials are viewed as unknown, the four-vectors $c_\mu$, $b_\mu$, and $d_\mu$ are unknown as well. Our goal is to check how they must be restricted to obtain lepton masses and physically acceptable solutions.

If the charged boson field $W_-=0$, then the equation \eqref{chFields:eq3272} splits into two Weyl-like equations which are similar to the conventional ones (except that $p_\mu$ is replaced by $p_\mu-b_\mu$ and $p_\mu- c_\mu$ for left- and right-handed parts of $\psi$ respectively). It is the last term in \eqref{chFields:eq3272} that makes the problem nontrivial. To see the connection between two parts of $\psi=(\chi, \xi)$, let us re-write the main equation in the spinor representation
\begin{equation}
	\begin{cases}\label{chFields:eq3274}
	(p-c)\bar\sigma\,\xi + d\bar\sigma\,\chi^C &=0\,,\\[1ex]
	(p-b)\sigma\,\chi + d\sigma\,\xi^C &=0\,,
	\end{cases}
\end{equation} 
where $d\sigma=d^\mu\sigma^\mu$, $d\bar\sigma=d_\mu\sigma^\mu$, $\sigma^\mu$ are four Pauli matrices, and the conjugate spinors are defined as
\begin{align}
	&\chi^C= i\sigma_2\chi^*\,,		&\qquad \xi^C= i\sigma_2\xi^*\,.
\end{align}
In the regular Dirac equation, it is the mass term that couples two spinors. In the extended equation \eqref{chFields:eq3272}, it is the term with charged boson field $W_-$. 
\\

\noindent\textbf{Rest frame: } The rest frame is defined by setting $\va{p}=0$ and choosing one of two spin-polarized forms \eqref{chFields:eq3342} for $\psi$. By selecting $s_z=+1/2$ for now, we set $\chi_2=\xi_1=0$. The system of linear equations for two remaining components $\chi_1$ and $\xi_2$ splits into two subsystems. The first one depends only on components $c_{1/2}$, $b_{1/2}$, and $d_{1/2}$ of vector coefficients; it does not include $p_0$ at all. In the rest frame with the spin projection aligned along the $z$-axis, the self-induced potentials (thus the coefficients $c_\mu$, $b_\mu$, and $d_\mu$) are expected to have the very specific form \eqref{chFields:eq3345}. Therefore, the first subsystem turns zero in such a rest frame. Instead, the second subsystem is nontrivial and has the following determinant
\begin{equation}\label{chFields:eq3205}
	\sqrt{\det M_0} =(b_0+b_3-p_0) (c_0+c_3-p_0)+ \abs{d_0+d_3}^2\,,
\end{equation}
where $M_0$ is the matrix $M$ evaluated in the rest frame. This expression is valid for both spin projections or both spinor forms \eqref{chFields:eq3342}. The eigenvalues depend only on the time- and $z$-components of self-induced potentials, see the discussion in subsection \ref{subSecLepton}. Two values of $p_0$ are given as 
\begin{equation}\label{chFields:eq3347}
	2 p_0 = b_0 + b_3 + c_0 + c_3 \pm \sqrt{(b_0+b_3-c_0-c_3)^2-4\abs{d_0+d_3}^2}\,.
\end{equation}
One eigenvalue turns zero ($m_1=0$) if
\begin{equation}\label{chFields:eq3346}
	(b_0+b_3)(c_0+c_3)= -\abs{d_0+d_3}^2\,,
\end{equation}
in which case the second eigenvalue becomes
\begin{equation}\label{chFields:eq3446}
	m_2= b_0+b_3+c_0+c_3\,.
\end{equation}
We will deal with small neutrino masses shortly; the constraint \eqref{chFields:eq3346} is satisfied only approximately in such a case. 

One can immediately see the seesaw-like relation in \eqref{chFields:eq3346} which makes one eigenvalue large if we force the second one to be small. First however, we have to show how these eigenvalues can be associated with effective masses of freely-propagating lepton states. In any way, the relation \eqref{chFields:eq3346} naturally appears within the proposed formalism. One should not be overly concerned with the explicitly non-covariant form of expressions \eqref{chFields:eq3205} - \eqref{chFields:eq3446}. They are derived from the equation $\det M=0$ which is Lorentz-invariant; ditto for its roots. Having evaluated them in one inertial frame (the rest one, for example), the invariant eigenvalues are the same for all other inertial frames.  

The solutions for both eigenvalues are given as 
\begin{align}
	&\psi_1 = N_1\smqty(\chi_1\\0\\0\\ -\frac{d_0+d_3}{c_0+c_3} \chi^*_1)\,,&\qquad \psi_2 = N_2\smqty(\chi_1\\0\\0\\ \frac{d_0+d_3}{b_0+b_3} \chi^*_1)\,,
\end{align}
where the normalization constants $N_{1/2}$ are not specified yet. Following our previous discussion, we expect that the magnitude of upper spinor must be much smaller (larger) than the magnitude of lower spinor for the neutrino-like (charged lepton-like) mode. If we request that
\begin{equation}
	\abs{c_0+c_3}\ll \abs{d_0+d_3}\ll\abs{b_0+b_3}\,,
\end{equation}
then the first and second solutions can be associated with the neutrino-like and charged lepton-like states respectively. Since the second eigenvalue which is associated with electron mass is positive, it immediately follows that
\begin{align}
	&b_0+b_3 > 0\,,				\qquad c_0+c_3<0\,.					
\end{align}
Remembering the definitions \eqref{chFields:eq3273} and speaking in relative terms, one can say the following about the self-potentials $A^{self}$ and $Z^{self}$. The combination of components ($A_0+A_3$) is large positive, while ($Z_0+Z_3$) is small negative. 

To introduce a nonzero neutrino mass, we have to expand the square root in \eqref{chFields:eq3347} over two small parameters $(c_0+c_3)/(b_0+b_3)$ and $\abs{d_0+d_3}/(b_0+b_3)$ which leads to 
\begin{gather}
\begin{aligned}
	&m_1 = c_0+c_3 + \frac{\abs{d_0+d_3}^2}{b_0+b_3}+\order{\dots}\,,\\[1ex]
	&m_2 = b_0+b_3 - \frac{\abs{d_0+d_3}^2}{b_0+b_3}+\order{\dots}\,.
\end{aligned}
\end{gather}
Therefore, the neutrino-like mass ($m_1$) is determined by the self-induced potentials $Z^{self}$ and $W^{self}_{-}$, as expected. The electromagnetic self-potential $A^{self}$ is the largest one among all four EW fields. Using the definitions \eqref{chFields:eq3273} and assuming that the time components are much larger than the $z$-ones in the rest frame under consideration, the above expressions are rewritten as
\begin{gather}
\begin{aligned}
	&m_\nu =\frac{g}{2\cos\theta}Z_0 + \frac{g}{4 \sin\theta}\frac{\abs{W_0}^2}{A_0}\,,\\[1ex]
	&m_e = e A_0 +\frac{g\cos2\theta}{2\cos\theta} Z_0- \frac{g}{4 \sin\theta}\frac{\abs{W_0}^2}{A_0}\,,
\end{aligned}
\end{gather}
where we also dropped the superscript from potentials to reduce clutter. The above expressions give the mass eigenvalues for single lepton family by means of the EW self-potentials. Two terms in the expressions for $m_\nu$ are probably close in magnitude, though the second term must be larger than the first one, since $(c_0+c_3)<0$. The relative smallness of first mass eigenvalue might also come from the fact that two terms in the expression for $m_\nu$ have signs opposite to each other. Instead, the first term in the expression for $m_e$ is much larger than the last two ones. Even if both neutral $Z$ and charged $W$ self-interactions contribute, the electromagnetic self-interaction is still the dominant contribution into $m_e$.

\section{Summary}
We derived the equation \eqref{chFields:eq3344} which extends the Dirac- and Weyl-like equations to the EW case. It is given for the quantity $\psi$ which represents one weak lepton iso-doublet (two Weyl spinors). We have also found its solutions in the form of plane waves under certain assumptions. Two types of solutions are distinguished by different values of mass eigenvalues; the connection to the conventional case, which corresponds to zero eigenvalues, is also given. We named these two types of new solutions as neutrino-like and charged lepton-like, since it remains open if they can represent real massive neutrinos and charged leptons. Though it will not be addressed in this work (our focus has been on developing the alternative and strictly equivalent description of conventional EW approach to the left-handed lepton sector), some additional considerations are given in the Appendix.

The equations \eqref{chFields:eq3344}  and  \eqref{chFields:eq3272} do not have a mass term in the conventional sense, however the scale of phase momentum  (energy-momentum vector $p_\mu$) is clearly set by the other vector coefficients
\begin{equation}
	[p_\mu]\sim [A^{self}_\mu]\sim [Z^{self}_\mu] \sim [W^{self}_\mu]\,,
\end{equation}
which are present in the main equation.  Physically, it means that the mass value (its bare value) is determined by the interaction of free-moving lepton-like state with its EW self-potentials. The structure of self-interaction terms is given by the regular EW terms, however the magnitudes of these self-potentials are free parameters. 

The model we developed here is phenomenological, since it does not predict the values of these self-potentials, or why there are three lepton generations for which extensions of relevant gauge groups might be required \cite{frampton_l_mu-l_tau_2020}. However, the model does show how to define neutrino-like masses alternatively to the Dirac and Majorana models. If one insists on a scalar mass term, then the theory has to deal with either the inert right-handed sector or the nonconservation of lepton numbers \cite{witten_lepton_2001, king_neutrino_2008}. Instead, the mass eigenvalues in our model is determined by well-known lepton-boson interaction terms which however are not scalar. 

The proposed framework can be straightforwardly meshed with Higgs mechanisms by simply adding the right-handed Lagrangian part, since the left-handed Lagrangian has not been  changed. Both mechanisms are truly complementary to each other. For example, the proposed mechanism can explain the bare masses in the lowest part of mass spectrum, while heavier fermion masses (which are closer to the Higgs mass) can be determined by the coupling to Higgs. It suggests a possible solution of why the coupling to scalar Higgs field does not lead to unrealistically high energy densities (which happens if everything couples to Higgs). 

It is also possible to extend the proposed model to higher order SMEFT terms by adding new interaction terms to the motion equations \eqref{chFields:eq3042}. Similarly, the model is straightforwardly extendable to the quark sector which will be tackled in a future work.  

Remarkably, we also managed to derive the analog of the seesaw relation, see the expression \eqref{chFields:eq3346}. Since the combination $b_0+b_3\sim m_e$ and $c_0+c_3\sim - m_\nu$ are shown to be measures of the electron and neutrino masses respectively, expression \eqref{chFields:eq3346} can be re-written as 
\begin{equation}\label{chFields:eq3348}
	m_e \, m_\nu \sim \abs{W_0^{self}}^2\,,
\end{equation}
where again the $z$-component of self-induced field $W_-$ is neglected. Therefore, the scale of product $m_e m_\nu$ is the magnitude of self-induced charged field squared.

We managed to derive the seesaw-like relation between the lepton-like masses and self-potentials in the rest frame. However, no values (bare values) of masses were possible to find without some additional input.  Obtaining the second relation, which connects the strength of these potentials with fermion field amplitudes  will allow finding the lepton-like masses in the proposed framework. Following footsteps of the conventional theory, other next tasks would be to apply the framework to quarks and bosons. At this point, we do not see any fundamental obstacles for extending the proposed model. However, any of these extensions is a big task in itself that clearly takes us outside the boundaries of this manuscript.

{\color{gray}\small
\appendix
\section{Additional considerations}
We derived two types of new plane waves that are solutions of equation \eqref{chFields:eq3344} which extends the Dirac- and Weyl-like equations to the EW case. The framework we develop is strictly equivalent to the conventional description of EW left-handed lepton sector. The solutions were however derived under certain additional assumptions which lead to nonzero mass eigenvalues for both types of solutions. In the proposed new representation, there also exists the direct link between the chirality polarization (ratio between upper and bottom spinors) and the electric charge of these lepton-like states. 

The question whether these new types of plane waves can represent real neutrinos and charged leptons is open \footnote{\color{gray} For this reason, we call these new solutions as neutrino-like and charge lepton-like states.} and will not be addressed in this work. Here however, we can share some additional considerations pertinent to this topic.

In the electroweak theory, the fermion masses are generated by coupling to scalar fields. The question regarding lepton  masses is effectively replaced by the question of why a given lepton type has a specific Yukawa coupling strength to the Higgs field \cite{quigg_unanswered_2009}. Since the Higgs couplings of leptons vary by twelve orders of magnitude, the question is whether some other mechanisms  of mass generation exist. The idea of radiative fermion masses was recently reviewed in \cite{weinberg_models_2020}. In such models, the heaviest fermions still receive their mass by coupling to the Higgs at tree level. Lighter fermions, however, acquire their masses in higher-order loops with virtual heavy particles. Taking into account that amplitudes of high-order loops are severely suppressed, this mechanism might potentially explain the vastness of mass spectrum. In any way, nearly all of numerous mechanisms of lepton mass generation, proposed in the literature, rely on coupling to scalar fields, even in the multidimensional extensions of SM \cite{de_anda_quark_2021}. Is it possible to introduce an alternative (and complementary) way of generating masses which would not contradict the well established results? Even if the Higgs is the key and proven part of the mainstream theory, it does not mean that there is no additional mechanism that can contribute into the masses of lightest leptons, which are many orders of magnitudes lighter than the Higgs. It is still the very active research area where different modifications of the original Higgs mechanism are proposed and investigated, see examples of recent works \cite{mccullough_implications_2021, englert_phenomenology_2020, anisha_extended_2021, cacciapaglia_review_2022, hinata_mass_2020, chaber_lepton_2018}. 
 Analogously, this new representation can be leveraged into a consistent extension to the conventional theory, without contradicting the existing, well proven results.

The biggest challenge for an alternative mechanism is how to introduce mass scales into field equations in a covariant and non-controversial way, similar to the Higgs mechanism. In the latter one, the mass terms are scalar, thus they must be of Dirac or Majorana type with their own sets of challenges \cite{witten_lepton_2001, zralek_50_2010, king_neutrino_2008, kim_inferring_2022}. The framework that is proposed in this study allows generating effective masses without scalar terms. Hence it avoids the necessity of choosing between the Dirac and Majorana models of massive neutrinos.   It might potentially open new possibilities in search for new physics beyond SM, in addition to what is already been actively discussed, see selected examples in  \cite{arguelles_new_2020, dune_collaboration_snowmass_2022,atkinson_muone_2022, baryshevsky_pseudoscalar_2020, baryshevsky_predicting_2022, dorsner_towards_2020}.  Additionally, the amount of literature on neutrinos-related topics is immense and will not be reviewed here; however, we would like to highlight several recent reviews \cite[p. 285]{zyla_review_2020-1}, \cite{dev_neutrino_2019}, \cite{altmannshofer_snowmass_2022}  and references therein.
\vspace{2ex}

\noindent\textbf{Relation to Higgs and symmetry breaking:}
It is important to clarify the relation of proposed mechanism to the Higgs and the symmetry breaking that leads to the mass generation in the conventional approach. The relevant terms of electroweak Lagrangian are given as
\begin{equation}\label{chFields:eq3026}
	\mathcal{L}_{ew} = i\bar{\psi}_L\gamma^\mu\partial_\mu\psi_L - f (\lambda+\chi)\bar{\psi}_R\psi_L - e A_\mu  \,\sum_l \bar{\psi}_L\gamma^\mu \psi_L+\dots\,,
\end{equation}
where $\psi_L$ and $\psi_R$ are left- and right-handed leptons, $f$ is the coupling strength to Higgs field $\chi$ which vacuum value is $\lambda$, and $A_\mu$ is the electromagnetic (EM) field.  The shown terms are the kinetic energy of left-handed lepton field $\psi_L$, the lepton-Higgs interaction term, and the lepton-EM interaction one correspondingly. The dots represent the corresponding terms for right-handed states, neutrinos, charged and neutral bosons, and Higgs. They are not shown explicitly in \eqref{chFields:eq3026}, since their roles remain unchanged in the proposed approach. 

The second term in \eqref{chFields:eq3026}, which includes the right-handed states of leptons, breaks the SU2 symmetry of electroweak Lagrangian. As a result of this symmetry breaking and the chosen  Higgs vacuum state,  particles (leptons, electroweak bosons $W_\mu$, $Z_\mu$, and quarks) all acquire masses while photons remain massless. Next, taking into account the radiative corrections into the self-energy, the effective mass of a lepton $m_l$ can be given as
\begin{equation}\label{chFields:eq5001}
	m_l = m_H + m_{A}^{rad}+\dots\,,
\end{equation}
where $m_H=f \lambda$ is the mass due to Higgs coupling and $m_{A}^{rad}$ is the contribution into lepton self-energy by virtual photons which are represented by field $A_\mu$. The dots represent radiative corrections generated by interactions with other types of virtual particles. We are still squarely within the conventional framework in which the techniques for evaluating these radiative corrections are well established however complicated they might be, especially in the case of hadronic vacuum polarization.

How can an additional effective mass appear in the equations of motion without violating the translation invariance and Lorentz covariance? Briefly, it can be outlined as follows. Let us assume that the potential $A_\mu$ has three parts
\begin{equation}\label{chFields:eq5000}
	A_\mu = A_\mu^{ext} +\partial_\mu \chi + A_\mu^{self}\,,
\end{equation}
where $\chi$ is an arbitrary scalar function (gauge), $A^{ext}_\mu$ is an external field while  $A_\mu^{self}$ is the field of free-moving charged particle. In the rest frame, it would be given by the Coulomb or Uehling potential $\phi(r)$ which is the former one  modified by the vacuum polarization around bare charges. We will consider only free-moving charges in this work, for which case $A_\mu^{ext}$ is  zero. Even after the renormalization, the part $A_\mu^{self}\sim \phi(r)$ remains infinite at the location of bare charge ($r\to 0$)  while the contributions of virtual photons into the self-energy (which depend on the regularization scale) are captured by the radiative corrections.

This infinite value of static potential $A_\mu^{self}\sim \phi(r\to 0)$ at the location of bare charge is largely ignored in the conventional field theory, at least in regard to particle masses, even if it has been frequently  speculated in the classical case that it should be connected to masses of charged particles (the electromagnetic origin of mass).  In this work, we develop the phenomenological approach within which a more fundamental theory of electroweak interactions can be applied to predict experimental results. We then assume that such a theory would lead to finite values of electroweak potentials  ($A_0$, $Z_0$, and $W_0$) at location of bare charge. In the approximation of free-propagating plane waves that we use here, they are just a set of numbers, with no space-time dependence.  We will not speculate here what form such a theory might have; all we need for our phenomenological approach to proceed is the assumption that  finite values of four electroweak potentials at bare charge origin exist. 

Next, if we are allowed assuming that the electroweak potentials of free-moving leptons, including the EM one $A^{self}_\mu\sim \phi(r\to 0)$ for example, are finite at charge origin, we can derive several non-trivial relations between masses of charged leptons and their neutrinos by using the conventional electroweak Lagrangian. Since the terms that describe the lepton-Higgs coupling and lepton-EM interaction enter the Lagrangian independently of each other, their contributions into the effective mass are also additive, in which case, the mass equation \eqref{chFields:eq5001} changes as 
\begin{equation}\label{chFields:eq5002}
	m_l^\prime = m_H + m_{A}^{rad}+m_{A}^{fin}+\dots\,,
\end{equation}
here the first and second terms  in \eqref{chFields:eq5002}  keeps its original meaning as in \eqref{chFields:eq5001}.  The third term gives the additional mass due to the finite self-potential $A_\mu^{self}$ which is not describable by the Maxwell physics. This is the key detail, since otherwise one would be able to immediately object to \eqref{chFields:eq5002} by pointing that the term $m_{A}^{rad}$ already describes the contributions due to the electromagnetic field. However, the conventional theory calculates radiative corrections by using the propagators which are solutions of Maxwell or Klein-Gordon equations which do not lead to finite values of potentials at charge origin. Therefore, assuming that such potentials are finite at $r\to 0$ necessarily imply the deviation from the Maxwell and Klein-Gordon physics (quantized or not). 

Since the contributions in \eqref{chFields:eq5002} due to Higgs and finite self-potentials are additive, the proposed mechanism is truly complementary to the Higgs one. It is possible that the proposed mechanism determines  masses of lightest leptons (that are much smaller than Higgs) which would make the Higgs couplings for heavier leptons constrained to a much narrow range. Assuming then that masses of lightest leptons are described by the proposed mechanism, we are able to show that the charged lepton mass $m_e\sim |A_0|$ is mostly given by the interaction with its EM field (but it is still not a purely electromagnetic one!) while the mass $m_\nu\sim |Z_0|$ of its corresponding neutrino is determined by the interaction with its neutral potential. Remarkably, the seesaw-like relation $m_e m_\nu \sim |W_0|^2$ between the masses of particles from  the same EW iso-doublet  also follows naturally from the applied framework. Hence, while the coupling strengths for a charged lepton and its neutrino are not connected in any way  in the Higgs-based model,  they are naturally connected by the seesaw relation in our approach.  Seesaw-like relations with neutrinos have been discussed before, see for example \cite{frampton_fermion_2009}, in the framework of gauged family unification.  An application of the proposed phenomenological model which would allow inferring additional information regarding the electroweak potentials of free-moving particles could also shed more light on possible deviation from the conventional theory to help address the known challenges with existing infinities. 
\vspace{2ex}

\noindent\textbf{Lepton-like states} 
The purpose of this manuscript is not to offer a complete or even a substantial partial solution of lepton mass problem, like a reduction in number of free parameters in SM (couplings to Higgs field, for example).  It would be an unrealistic expectation at this stage. Instead, our goals are much more modest. We attempt to expose a potential gateway from which a consistent extension of conventional EW theory, which does not contradict with any of well proven experimental support and theory constructs, can be found.

A massive particle must have both left- and right-handed components \cite{witten_lepton_2001}, as can be seen by considering its behavior under general boosts and rotations. Specifically, the Dirac plane wave has the left- and right-handed components of equal magnitude in the rest frame, see for example, (3.49) from \cite{peskin_introduction_1995}.   Such a state does not have any left-right asymmetry per se, opposite to what we see in the EW interactions (nearly 100$\%$ chirally polarized).  The mass of such a particle is a scalar m that is introduced into the Lagrangian by hand (or by using the Higgs field in a certain vacuum state).

In the new representation, a massive lepton-like state also has both left- and right-handed components. However, the left-right asymmetry is incorporated into the state description: a massive charged lepton-like state has a large left-handed component and very small right-handed one.  In return, an antilepton-like state has a large right-handed component and very small left-handed one.  To avoid any issues with the behavior under boosts and rotations, such states must be fully spin-polarized in their respective rest frames, see (41)  for a general boost and its specific version (42) for the boost in z-direction. As we showed, we can then obtain several scalars, which are interpreted as contributions into masses, for free-moving leptons from the EW currents for such states and finite self-potentials, without invoking the Higgs mechanism.

\vspace{2ex}

\noindent\textbf{Effective mass:} Conventionally, mass terms are introduced into field equations as products of constant $m$ with scalars made out of field amplitudes. For this reason, right-handed leptons are required in conventional Lagrangians to introduce scalar Dirac-like mass terms (and for anomaly cancellations which also require contributions from corresponding quarks \cite{frampton_l_mu-l_tau_2020}). It is truly remarkable that right-handed states of particles and left-handed ones of antiparticles do not interact with charged EW bosons.  Since the proposed framework allows generating non-zero values of four-momentum at rest without such scalar terms, it would be helpful to elaborate more on how it can be achieved.

How can effective masses appear in the equations of motion without violating the translation invariance and Lorentz covariance? Consider, for example,  a free-motion without external fields which is described as
\begin{equation}\label{chFields:eq3062}
	i\slashed\partial \psi - \underbrace{e \slashed A^{self}\psi -(\dots)^{self}}_{ S^{self}\psi}=0\,, 
\end{equation}
here $A^{self}_\mu$ is not yet known self potential, or the potential of free-moving charge.  The goal is to find whether the inclusion of  terms $S^{self}$  can be done in a consistent way. It would be impossible to generate masses in a consistent way with having only one gauge potential $A_\mu$. However, the EW theory with four gauge potentials provides a sufficiently advanced framework for obtaining non-trivial results, as we will see below. We also do not expect that it would be possible to reduce the self-interaction $S^{self}$ to some kind of Dirac or Majorana mass terms, since it will create well-known challenges. For example, a Dirac mass term requires right-handed states that do not interact with the charged EW fields (enabling the scalar mass term seems to be their only role in theory), while a Majorana term does not conserve either lepton number or charge.  Specifically, we show in this work that for free-moving leptons $\psi(x)=\psi e^{-ipx}$, the evolution equation reduces to a system of algebraic equations for components of $\psi$
\begin{equation}
	i\slashed\partial \psi - S^{self} \psi=0\qquad \to \qquad  \slashed p\psi - S^{self}  \psi=0
\end{equation}
where the operator $S^{self} $ is not a scalar one. The mass eigenvalue is then obtained by squaring the momentum vector: $p_\mu p^\mu = m^2$. The proposed mechanism could be used as a complement to the Higgs one to address the question why Higgs coupling constants span such a wide range.

The outlined task is different from the known approaches, either the Dirac-Maxwell system or self-energy evaluations in the QFT.  The latter one is the standard step which requires multi-loop calculations followed by renormalization. It is well known that fermion masses run for this reason; hence different values are given at different energy scales. For example, the standard procedure takes into account multiple acts of emission and re-absorption of virtual photons which however are assumed to propagate as solutions of Maxwell equation. Instead, we do not assume that $A^{self}_\mu$ obeys the Maxwell equation; effectively, we attempt to evaluate the self-induced mass in the \emph{zero} order by relaxing requirements that are not strictly proven. Then, the mass value can be refined in next orders by standard QFT procedures to take into account the residual self-energy as the radiation correction. Clearly, it is the hypothesis that must be proven by obtaining consistent results. 

}

\bibliographystyle{elsarticle-num-names}

\end{document}